# DensePANet: An improved generative adversarial network for photoacoustic tomography image reconstruction from sparse data

**Hesam Hakimnejad**

Shiraz University, Department of Computer Science Engineering and Information Technology, Shiraz, Iran
hesam.hakimnejad@shirazu.ac.ir

**Zohreh Azimifar**

Shiraz University, Department of Computer Science Engineering and Information Technology, Shiraz, Iran
azimifar@uwaterloo.ca

**Narjes Goshtasbi**

Shiraz University, Department of Computer Science Engineering and Information Technology, Shiraz, Iran
nargesgoshtasbi.ng@gmail.com

**Abstract**

Image reconstruction is an essential step of every medical imaging method, including Photoacoustic Tomography (PAT), which is a promising modality of imaging, that unites the benefits of both ultrasound and optical imaging methods. Reconstruction of PAT images using conventional methods results in rough artifacts, especially when applied directly to sparse PAT data. In recent years, generative adversarial networks (GANs) have shown a powerful performance in image generation as well as translation, rendering them a smart choice to be applied to reconstruction tasks. In this study, we proposed an end-to-end method called DensePANet to solve the problem of PAT image reconstruction from sparse data. The proposed model employs a novel modification of UNet in its generator, called FD-UNet++, which considerably improves the reconstruction performance. We evaluated the method on various *in-vivo* and simulated datasets. Quantitative and qualitative results show the better performance of our model over other prevalent deep learning techniques.

1. **Introduction**

Photoacoustic tomography (PAT) also named optoacoustic tomography (OAT) is a hybrid non-ionizing biomedical imaging technology based on the photoacoustic effect [1]. The photoacoustic effect refers to the propagation of acoustic waves as a result of irradiation from tissue chromophores by a short-pulsed laser. Ultrasound imaging offers superior spatial resolution compared to optical imaging, primarily due to reduced scattering in biological tissues. In contrast, optical imaging delivers enhanced contrast images, primarily linked to factors like oxygen saturation and hemoglobin concentration in biological tissues [2]. PAT allows us to benefit from these two important features, together [3]. This multi-scale imaging modality is a useful tool for imaging different structures in biological tissues anatomically, functionally, and molecularly for diagnostic and treatment purposes [4]. PAT has been effectively utilized for imaging purposes across a diverse array of applications, including the imaging of vasculature, tumors, skin cancer, mouse brain vasculature [5], imaging the brain and whole body of small animals [6] and breast cancer diagnosis [7]. The process involves using ultrasound transducers to capture laser-induced photoacoustic waves, which are then used to form an initial pressure image. Reconstruction algorithms are subsequently applied to reconstruct the final images. Filtered back-projection (FBP) [8], time reversal (TR) [9], and delay-and-sum [10] are the most common standard PAT image reconstruction methods.

For accurate reconstruction in PAT, it is essential to fully enclose the target with the transducer and have a sufficient density of transducer elements. However, practical constraints, such as incomplete spatial coverage or the need for faster data acquisition, often limit the transducer's access to a limited portion of the photoacoustic (PA) signals. As a result, images reconstructed using standard methods exhibit poor quality and prominent artifacts due to under-sampling [11]. This issue, known as the limited-view problem [12], makes a significant barrier to the analysis of medical images obtained from the reconstruction process. This situation mostly reduces image quality, leading to a loss of important information concerning the imaging target. To tackle the issue of the limited-view problem, iterative image reconstruction algorithms have been suggested, aiming to enhance the quality of images through noise and artifact reduction using signal sparsity and low rankness [13]. In spite of their strong performance, these methods have some drawbacks. The major disadvantages of these methods include increased computational complexity, parameter tuning, memory requirements, non-uniqueness and limited applicability [14]. During recent years, a number of studies have attempted to solve the limited-view problem in PAT image reconstruction algorithms by utilizing deep learning (DL) methods [15]. DL-based non-iterative PAT image reconstruction methods can be categorized into two types based on the input data: direct DL methods and DL as post-processing methods. In the latter, conventional methods like time reversal are initially used to perform a primary reconstruction of the measurement data. Subsequently, DL techniques are employed to eliminate artifacts and enhance the quality of the reconstructed images. direct DL methods, as their name implies, directly convert the measurement data into reconstructed images without an intermediate primary reconstruction step [16].

Convolutional neural networks (CNNs) are famous DL methods that have shown remarkable advancements in various image processing and computer vision tasks, like semantic segmentation, image classification, object detection, and localization [17] to name a few. This progress has been extended to the domain of biomedical image analysis, and in particular, significant improvements have been achieved in the reconstruction of medical images by leveraging CNNs.

In 2019, Antholzer et al. [18] proposed a direct and efficient reconstruction method based on DL. They used a deep CNN to perform image reconstruction, combining the PAT filtered back-projection algorithm with the UNet architecture. This approach allows for image reconstruction without the need for time-consuming forward and adjoint problem solutions. Their numerical results demonstrate that this method generates high-quality images from sparse data that are equivalent to cutting-edge iterative techniques. Hauptmann et al. [19] developed a deep neural network that produces 3D images with high resolution using only a few photoacoustic measurements. They separated the computation of gradient information from training to deal with the complexity of the photoacoustic forward operator. Additionally, a prior knowledge of desired image structures was learned during the training process.

On the other hand, UNet is considered a breakthrough in the field of biomedical image analysis, and it has been applied to many related tasks with a significant success rate. Waibel et al. [16] introduced a technique that utilizes DL to reconstruct the initial pressure from a limited view of PA data. Their approach involves using a UNet-like architecture, which is trained using pixel-wise regression loss on the PA images. Similarly, Davoudi et al. [20] employed the UNet structure to eliminate artifacts from reconstructed images. Guan et al. [11] suggested a revised CNN structure called fully dense UNet (FD-UNet) to eliminate artifacts from 2D PAT images. Additionally, they highlighted the significance of training deep neural networks with top-notch real images. Later, generative artificial intelligence also entered the game. For instance, a GAN based solution was proposed by Vu et al. [21] to decrease limited-bandwidth and limited-view artifacts from photoacoustic computed tomography (PACT) images.

Being inspired by the impressive potential of GANs in generating realistic images [22], their success in paired image-to-image translation tasks [23], and their application in unpaired image-to-image translation ones [24] we developed a supervised GAN-based post-processing reconstruction algorithm called DensePANet for reconstructing PAT images with high quality. We built a novel architecture called FD-UNet++, combining the features of dense blocks and UNet++, as the generator of this model.

The remainder of the paper is structured as follows: Section 2 reviews the theoretical background and Section 3 presents the methodology of our work. Details of datasets and experimental results are detailed in Section 4 and in the end, we conclude the paper in Section 5.

## 2. Background Review

Before delving into the proposed model architecture, it is inevitable to describe the fundamentals of UNets as well as GANs, since our model is mainly based on these networks.

### 2.1. UNet-based Reconstruction

UNet has emerged as an effective CNN architecture, demonstrating notable advantages, such as efficiency in handling a limited amount of training data, rendering it suitable for medical applications [25]. The UNet architecture is composed of two essential components: a contracting path (encoder) and an expanding path (decoder). This encoder-decoder-based CNN framework employs skip connections between the encoder and decoder paths, which has found extensive utility in various medical imaging applications, particularly in the context of segmentation tasks [26]. UNets leverage these skip connections along with the encoder-decoder structure, to effectively learn latent translations between input and output representations. The encoder employs convolutional and pooling layers to down-sample the input image, resulting in a lower-resolution representation. Conversely, the decoder employs deconvolution layers whose numbers match the number of layers in the encoder. This design ensures symmetry between the encoding and decoding processes, allowing the feature maps to be reconstructed to their original dimensions. In other words, for every downsampling operation performed in the encoder, an equivalent up-sampling operation is performed in the decoder, maintaining consistency in the number of layers and enabling the restoration of the original image size. By incorporating down-sampling and up-sampling layers, UNet effectively learns features at various resolutions while maintaining computational efficiency [27].

Recently, UNet++ outperformed the standard UNet as a more robust and advanced architecture for biomedical image segmentation [28]. In UNet++ a significant modification to the original UNet architecture is introduced, i.e., redesigned and dense skip pathways. To illustrate more, sub-networks are connected via a series of nested, dense skip connections instead of plain ones. This idea reduces the semantic differences between encoder and decoder paths, and thus, results in better segmentation results [29].

Deep CNNs are widely employed in order to extract desirable features from images. The depth of a CNN assumes a crucial role in its capacity to capture intricate features from images. However, as the network depth increases, the issue of gradient vanishing becomes prominent. Gradient vanishing refers to the diminishing gradient signals during back-propagation, which can impede the convergence and stability of the model during training. It is vital to adopt strategies to mitigate this phenomenon and ensure successful model optimization [30]. Fully dense UNet, a. k. a. FD-UNet, is a more effective architecture compared to UNet in artifact removal [11]. In this variant of UNet, each pathway integrates dense blocks, as introduced in the DenseNet [31]. These dense blocks consist of four densely connected layers, facilitating enhanced feature extraction and representation within the FD-UNet architecture. This modification contributes to the overall effectiveness and performance of the FD-UNet model, Fig 1. The input of each layer is taken from a channel wise concatenation of outputs from the previous layers. This enables the network to reuse learned features in dense blocks, which enhances the representational capacity of the network. By leveraging previously learned features, the network eliminates the need for redundant feature learning while promoting the acquisition of diverse feature sets. In this architecture, the number of feature maps follows an exponentially increasing pattern, starting with 64 feature maps for the first block, 128 feature maps for the second, and so on. This pattern continues throughout the expanding path, which is also composed of four blocks.

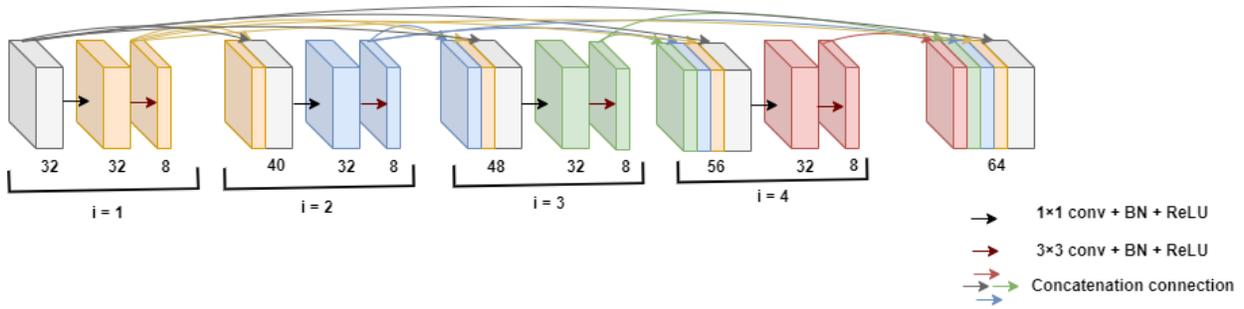

Fig 1. Four layered dense blocks. The input for the following layers is created by concatenating the feature maps from earlier layers.

Overall, UNet-based image reconstruction algorithms have demonstrated significant potential in advancing the area of medical image reconstruction. With further research and development, these techniques are expected to continue evolving and contributing to advancements in various domains that rely on accurate image reconstructions.

### 2.2. GAN-based Image Reconstruction

The performance of UNet-based models is sometimes limited by blurry results because the loss function reduces the Euclidean distance between predictions and the target[21]. Generative adversarial networks (GANs) address this issue by simultaneously training a discriminator network to differentiate between real and generated images. The introduction of GANs has been a significant breakthrough in DL, revolutionizing multiple domains by generating realistic and high-quality synthetic data through adversarial training. The innovative framework of GANs has sparked tremendous advancements in areas such as image synthesis, data generation, and unsupervised representation learning. GANs have witnessed widespread adoption for image generation and augmentation, leading advancements in various fields, including medicine.

A GAN has two components, a discriminator and a generator, Fig. 2. The generator learns to sample from the data distribution and to synthesize new instances. The discriminator, on the other hand, is used to distinguish between synthesized and real samples, competing with the generator. This adversarial training process facilitates the unsupervised learning of hierarchical representations within both networks. By providing random noise as input to the generator, GANs can generate realistic images that resemble various entities, including non-existent people and a wide range of other objects or scenes.

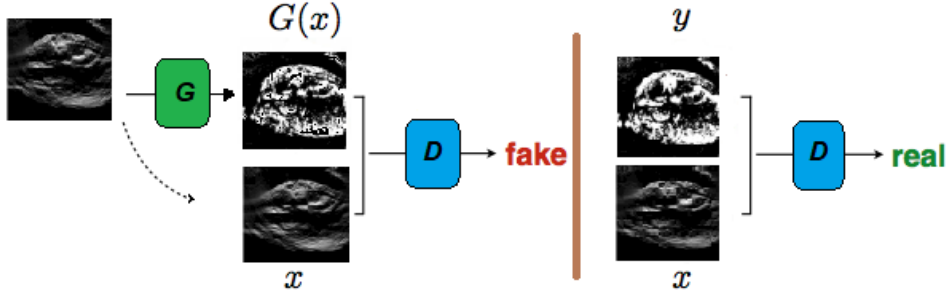

Fig 2. Artifact elimination in PAT imaging with undersampled data using a generative adversarial network approach based on the FD-UNet++ architecture.

It is important to highlight that during training, the discriminator model is updated directly, while the generator model is updated based on feedback from the discriminator. This simultaneous training process creates an adversarial dynamic, where the generator aims to deceive the discriminator, while the discriminator strives to accurately distinguish between genuine and synthetic images. Through this iterative process, both models can improve their performance over time. The generator receives random noise as input, while the discriminator receives both actual data and fake data produced by the generator as inputs. A GAN's objective function is described as follows:

$$\min_{G} \max_{D} V(D, G) = E_{x \sim p_{data}(x)}[\log D(x)] + E_{z \sim p_{z}(z)}[\log(1 - D(G(z)))] \quad (1)$$

where D is the discriminator, G is the generator, z is the input noise vector, and x is the real image. A variant of GANs known as conditional GAN (cGAN) [32] introduces the concept of conditional generation, where the generator model produces images based on specific conditions or inputs. Unlike traditional GANs that generate images solely from random vectors, cGANs incorporate additional information or constraints to guide the image generation process. By conditioning the generator on specific inputs, such as class labels or other auxiliary information, cGANs allow the generation of images with desired characteristics or attributes. This approach expands the capabilities of GANs, opening up new possibilities in image analysis. In particular, the Pix2Pix model is a popular cGAN in image-to-image translation, introduced by [23]. The generator in the Pix2Pix model is built upon the UNet architecture, making it particularly suitable for image reconstruction, and the discriminator follows the standard approach of deep CNNs. The objective function of cGAN can be stated as:

$$L_{cGAN}(D, G) = E_{x,y}[\log D(x, y)] + E_{x,z}[\log(1 - D(x, G(x, z)))] \quad (2)$$

where z indicates the noise vector, x is the input (artifactual) image and y stands for the ground truth (artifact-free image).

Pix2Pix objective function is obtained by Combining the GAN objective with traditional loss functions, like the $l1$-norm. It offers advantages, thereby, the discriminator's role remains the same, but the generator's task expands to confuse the discriminator while producing outputs close to the ground truth. This integration promotes realistic and accurate output generation. The final objective function is as below:

$$Loss = arg \min_G \max_D L_{cGAN}(D, G) + \lambda L_{l_1} \quad (3)$$

## 3. Proposed Network Architecture

In this section, we provide an elaborate explanation of the primary elements constituting the suggested model, which we call DensePANet. Overall, we built DensePANet based on the Pix2Pix model, proposing the FD-UNet++ architecture as its generator, Fig. 3. For the generator, we developed a novel version of UNet that integrates features from DenseNet as well as UNet++. This is an encoder followed by a decoder similar to those of UNet. Instead of regular convolutional operations, we opted to use the dense blocks with layer-wise specific input-output feature map sizes. After each dense block, we used a 2 × 2 maxpooling to halve the feature map size to the subsequent layer, followed by increasing the number of channels by a factor of 2. A string of dense blocks and max-pooling operations builds the contracting path of the architecture and for the expanding path, we used a series of transpose convolutions and concatenations.

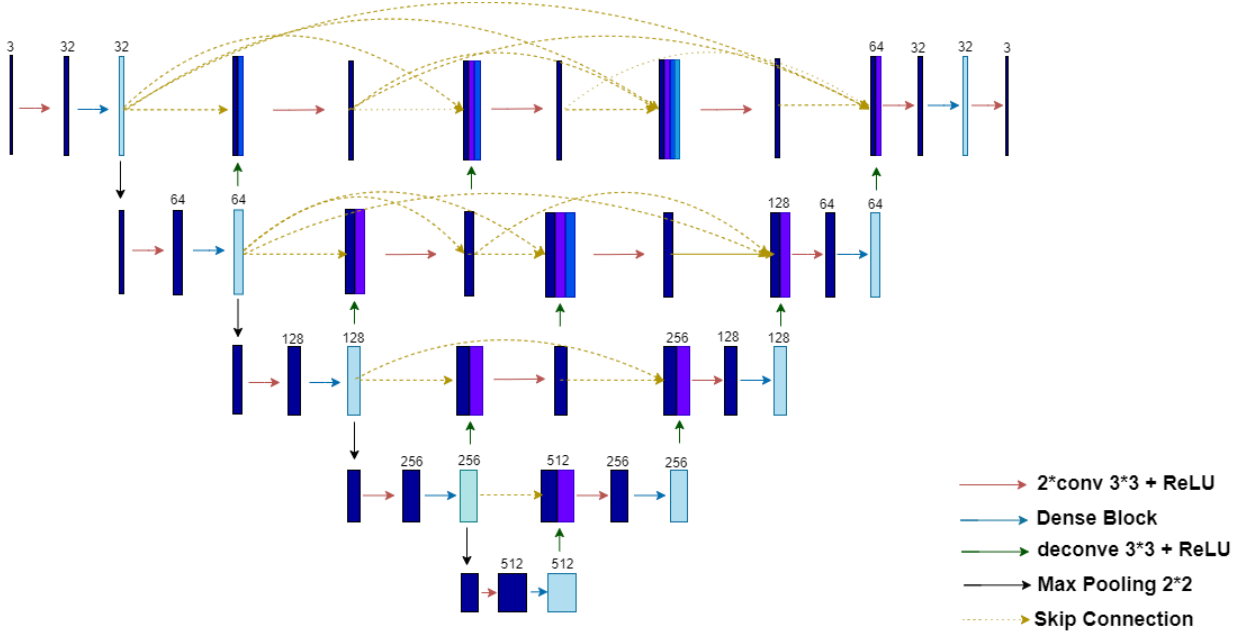

Fig 3. The proposed DensePANet's generator network architecture. It comprises of a convolutional encoder on the left side and a decoder based on the UNet++ model on the right side. The model is designed to integrate both low-level and high-level features during training, facilitating comprehensive feature representation.

Generally, skip connections play a substantial role in the information flow between expanding and contracting paths and allow the combination of dissimilar feature maps in encoder-decoder sub-networks. Therefore, it seems that redesigning skip connections, like what can be seen in UNet++, can exert a significant impact on how the model works. This design allows each node to have access to a comprehensive range of information for more effective feature extraction.

The following equation is the calculation of the stack of feature maps, denoted by $S^{m,n}$

$$S^{m,n} = \begin{cases} \odot(S^{m-1,n}), & n = 0 \\ \odot([[S^{m,k}]_{k=0}^{n-1}, US(S^{m+1,n+1})]), & n > 0 \end{cases} \quad (4)$$

where m stands for the downsampling layer along the encoder and n for the convolution layer of the dense block along the skip pathway, function $\Theta(.)$ means the convolution operation, [.] stands for the concatenation layer and $US(.)$ denotes an up-sampling layer.

Prior to each convolutional layer, a concatenation layer is applied. The concatenation layer combines the up-sampled output from the lower dense block with the output from the previous convolutional layer inside the same dense block. This fusion of features enables information exchange and enhances the connectivity between the nodes in the skip pathway. To illustrate more, Nodes situated at level $n = 0$ exclusively obtain a single input from the preceding layer of the encoder. Nodes positioned at $n = 1$ receive a pair of inputs, both stemming from the encoder's sub-network but originating from two successive levels. For nodes located at a level $n > 1$, they receive a total of $n + 1$ inputs. Among these inputs, $n$ inputs are sourced from the outputs of the previous $n$ nodes within the identical skip pathway. The last input is obtained from the output that has been up-sampled from the skip pathway at a lower level.

FD-UNet++ also employs dense blocks[11] to extract feature maps, each with a growth rate denoted as $k_l$ for a given spatial level, $l$. Initially, we define hyperparameters $k_1$ and $f_1$ as starting values. To enhance computational efficiency, the value of $k_l$ is adapted at every spatial level. In this implementation $k_l$ and $f_l$ are defined as follows:

$$k_l = 2^{l-1} \times k_1 \qquad (5)$$
$$f_l = 2^{l-1} \times f_1$$

The features extracted from the contracting path are combined with those originating from the up-sampled feature maps. This fusion ensures that the up-sampling process preserves higher-resolution features.

The $i^{th}$ layer within the dense block, as depicted in Figure 1, produces an output with $k_l$ feature maps and takes an input with $F + k_l \times (i - 1)$ feature maps. Here, $F$ represents the total number of feature maps in the initial input of the dense block. The process of feature learning involves a combination of 1×1 and 3×3 convolutions. These convolutions are followed with Batch Normalization (BN) and Rectified Linear Unit (ReLU) activation function. The inclusion of the 1 × 1 convolution enhances computational effectiveness by decreasing the input dimension to $F$ feature maps before applying the more intensive 3 × 3 convolution. Consequently, $k_l$ feature maps are learned from the downsized input through a 3 × 3 convolution.

The discriminator in our setup is a CNN, comprising of six convolutional layers, each followed by Instance Normalizations and ReLU activations. The first convolutional layer generates a feature map of 64 channels from the input image. Then, the number of channels doubles each time that the feature map undergoes a convolutional layer, with the exception for the fifth layer. The output tensor is calculated in the final layer by adjusting the number of channels to three. More importantly, we use a patch-based discriminator as is suggested in the Pix2Pix model, which produces predictions across particular regions of the input image rather of the entire image. This type of patch-level architecture not only has fewer parameters but also is capable of working with images of any size. This method can also outperform standard discriminator formulations as it imposes additional constraints that promote the presence of clear, high-frequency detail.

## 4. Experiments and Results

In this section, multiple experiments are conducted to validate the efficiency of the method we have proposed.

### 4.1. Datasets and Preprocessing

Obtaining ground truth is a key problem for supervised reconstruction methods as there is a lack of suitable public datasets, especially, in PAT images. On the other hand, DL-based methods need to be trained on large datasets to obtain accurate and robust models, making data augmentation, and in the case of PAT images, data simulation, an inevitable step in our work.

In order to assess the effectiveness of the suggested approach, we carried out extensive experiments on three datasets. The descriptions of these datasets are as follows. We used data augmentation, by means of cropping, zooming, and flipping the images horizontally or vertically.

More importantly, we simulated photoacoustic images from existing public non-PAT images to address the issue of limited public PAT dataset available. Table 1 shows a summary of our datasets.

Table 1 Details of Different Datasets in Our Experiments.

| Dataset | Original size | Augmented size | Size | Split (Train/Test) |
|---|---|---|---|---|
| Simulated Vessels | 80 | 2000 | 256 × 256 | 1600/400 |
| Mouse-Abdomen | 274 | 5000 | 256 × 256 | 4000/1000 |
| Brain Tumor MRI | 1321 | No Augmentation | 256 × 256 | 1057/264 |

### 4.1.1 Simulated Vessels

As PAT has found extensive application in the visualization of blood vessel arrangements, a synthetic model mimicking blood vessels with diverse initial pressure patterns is chosen. This model serves to showcase how various algorithms perform in the task of reconstructing these blood vessel structures. We employed the MATLAB toolbox k-Wave [33] to produce the synthetic PAT dataset. The detectors were uniformly positioned around the object at a radius of 45 mm. A sound speed of 1500 m/s was taken into account, and the area for reconstruction was defined as $10 \times 10\ mm2$. The initial pressure distribution was based on publicly available fundus oculi vessel data from the DRIVE database [34], comprising a total of 20 images. Each image was subdivided into four equal sized segments of $256 \times 256$ dimensions, and subsequently augmented to yield a total of 2000 images. This augmented dataset was then partitioned, with 1600 images designated for training and 400 images for testing. It's important to highlight that the k-Wave toolbox's time reversal method was employed to reconstruct the initial image from the simulated photoacoustic time series data.

### 4.1.1. Mouse-Abdomen

For the second experiment, we used the publicly available Whole-body mouse dataset [20]. We selected a set of data with 256 tomographic projections from this dataset. This involves 274 images, bringing the total to 5000 images through our augmentation process.

Davoudi et al. [20] acquired this dataset using a whole-body animal imaging scanner based on cross-sectional tomographic geometry. The detection process utilized a ring-shaped transducer comprising 512 cylindrically focused individual sensors. The cross-sectional images from the abdomen were reconstructed by employing all 512 tomographic projections (ultrasound sensors/detection elements) and were considered as artifact-free images or ground truth ones. In fact, the utilization of full-view geometry and a dense distribution of detection sensors guarantees sufficient sampling of acoustic signals throughout the whole target. This approach resulted in excellent image quality, especially when all 512 detected photoacoustic signals were used in the reconstruction process. We divided the resultant images into two sets, comprising 4000 images for training and 1000 images for testing.

### 4.1.2. Brain Tumor MRI

We also included a magnetic resonance imaging (MRI) brain dataset, publicly available from the Kaggle website [35]. This dataset contains 1321 images for classifying brain tumors, which we divided into 1057 images for training and 264 images for testing. These images all had dimensions of $256 \times 256$. In fact, we first converted these brain MRI images into raw photoacoustic signals using K-Wave Toolbox. Then, using a traditional reconstruction method such as time reversal method, we performed an initial reconstruction of these simulated acoustic signals. Now, it looks like we have simulated images in photoacoustic imaging modality. Finally, by using different methods based on deep learning, we improved these brain images and removed the artifacts that were created due to the problem of limited view.

### 4.2. Evaluation Metrics

The image reconstruction quality is assessed using the structural similarity index (SSIM) and peak signal-to-noise ratio (PSNR) by Utilizing the test data [36]. These two metrics are comprehensive criteria to evaluate the reconstruction ability of a network as PSNR presents an overall assessment of the quality of images while SSIM measures the similarity of patterns in pixel intensities across regions. The SSIM is a perceptual metric that quantifies the deterioration of image quality and offers a normalized mean value of structural similarity between the two pictures. Equations 7 and 8 indicate how these metrics are to be calculated.

$$SSIM(x,y) = \frac{(2\mu_x\mu_y + C1)(2\sigma_{xy} + C2)}{(\mu_x^2 + \mu_y^2 + C1)(\sigma_x^2 + \sigma_y^2 + C2)} \tag{6}$$

Here $\mu_x$ represents the average value of the pixels in x, while $\mu_y$ represents the average value of the pixels in $y$. The variances of x and $y$ are denoted as $\sigma_x^2$ and $\sigma_y^2$, respectively. The covariance between $x$ and $y$ is represented by $\sigma_{xy}$.

$$PSNR = \frac{10\log_{10}(peakval^2)}{MSE} \tag{7}$$

where the term *peakval* represents the maximum pixel value found in the image. The Mean Squared Error (MSE) is determined by averaging the squared differences between the reconstructed image and its ground truth; lower values denote better performance.

*4.3. Results*

On several datasets, we ran experiments to demonstrate how well the suggested model worked. Moreover, we implemented several strong models in the domain of image reconstruction, and assessed them using these datasets in order to compare fairly with the operation of the suggested architecture. The ability of our proposed method to rebuild images with sharp edges and few artifacts is quite evident.

On an NVIDIA GeForce 3060 GPU, we trained and assessed each model using Keras with a Tensorflow backend. For model optimization, we used the Adam optimizer [37] with a learning rate of 2e-4, momentum of 0.5, and batch size of 1.

Table 2 compares the performance of our proposed model to other approaches in terms of SSIM and PSNR for the Simulated Vessels, Mouse-Abdomen, and Brain Tumor MRI datasets. It is evident that our approach often outperforms any other approach in image reconstruction in terms of SSIM and PSNR.

Table 2 Comparison Between Reconstruction Performances of Previous Works and The Proposed Model Applied to Three Different Datasets.

| Dataset | | UNet | YNet | UNet++ | FD-UNet | DensePANet (ours) |
|---|---|---|---|---|---|---|
| Simulated Vessels | SSIM | 0.8308±0.1407 | 0.9322±0.0030 | 0.8599±0.0527 | 0.9473± 0.0139 | **0.9733±0.0032** |
| | PSNR | 20.6166±2.9360 | 21.6223±0.0119 | 19.8636 ±0.8785 | 23.7262 ± 0.3288 | **25.7718±0.5997** |
| Mouse-Abdomen | SSIM | 0.7509±0.0154 | 0.7800 ± 0.0039 | 0.7776± 0.0019 | 0.8064 ± 0.0138 | **0.8602±0.0064** |
| | PSNR | 17.7014±0.4864 | 17.6960± 0.0522 | 17.2020±0.3733 | 18.6775±0.0117 | **20.0831± 0.2912** |
| Brain Tumor MRI | SSIM | 0.7523±0.0242 | 0.7826±0.0017 | 0.7883±0.0033 | 0.8003±0.0032 | **0.8504± 0.0011** |
| | PSNR | 20.5583±0.4277 | 19.9501±0.0662 | 20.1228± 0.4602 | **26.6770±0.0169** | 22.0917±0.0710 |

Clearly, UNet yielded the worst performance among all models in all three experiments. The cause lies in the structure of the network, which consists exclusively of contraction and expansion blocks linked by straightforward skip connections between the encoder and decoder segments. Consequently, detailed information from the lower levels was overlooked in the expansion layers. The UNet++ approach outperformed the UNet due to the implementation of nested skip connections connecting the encoder blocks with their corresponding decoder blocks. The Ynet [38] framework is a CNN based architecture that merges two encoder networks with a single decoder pathway (so it is Y shaped), with the efficient utilization of information from both raw data and the beamformed image. YNet outperforms UNet and UNet++ in simulated vessels and mouse-abdomen datasets.

FD-UNet made considerable improvements in the reconstruction task, increasing SSIM and PSNR and finally, our method achieved the highest SSIM and PSNR except in one experiment.

More precisely, FDUNet reaches a better PSNR when it comes to Brain Tumor MRI images, while our proposed method achieves the best PSNR on Simulated Vessels and Mouse-Abdomen datasets by notable differences compared to FD-UNet.

Moreover, Figures 4, 5, and 6 depict the reconstructed images from different test datasets on some sample images using the proposed model and other approaches. It is clear that outputs from our proposed model have better qualities compared to previous methods and, assuming higher similarity to the ground-truth. Especially when compared to other methods, our approach reconstructs the targets with more accurate boundaries and less missed components.

Especially as seen in figures 5 and 6, the results from our proposed method are clearer and less blurry.

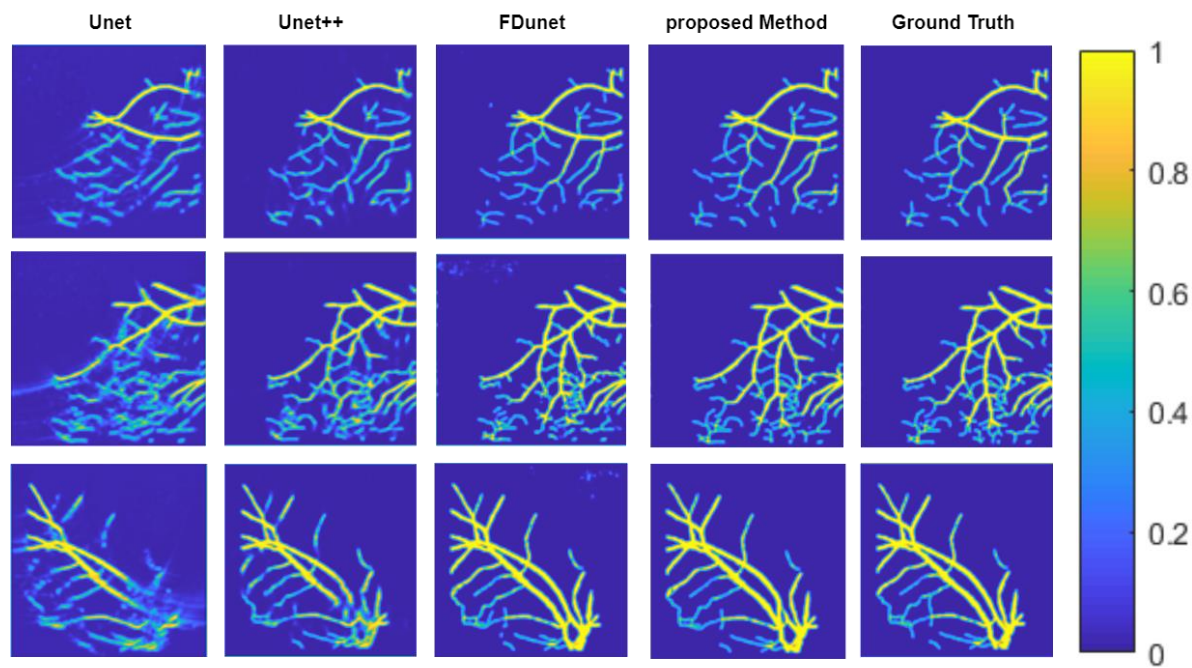

Fig 4. The performance of DensePANet applied to Simulated Vessels dataset. The pictures from left to right are the reconstructed images resulting from the models UNet, UNet++, FD-UNet, and DensePANet, and the ground truth.

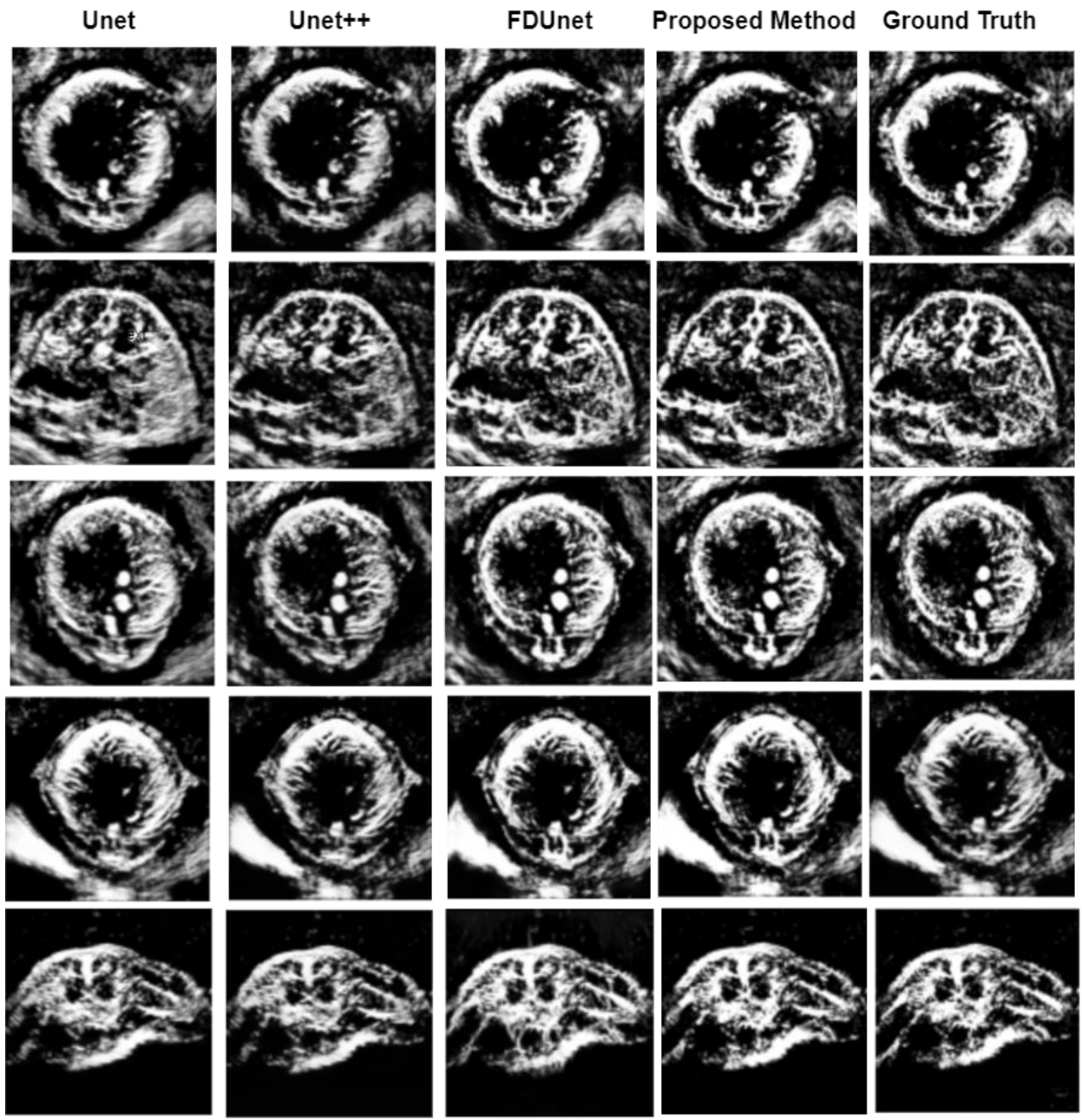

Fig 5. The Mouse-Abdomen dataset was used to evaluate the performance of DensePANet. The reconstructed images from the models UNet, UNet++, FD-UNet, and DensePANet, as well as the ground truth, are displayed from left to right.

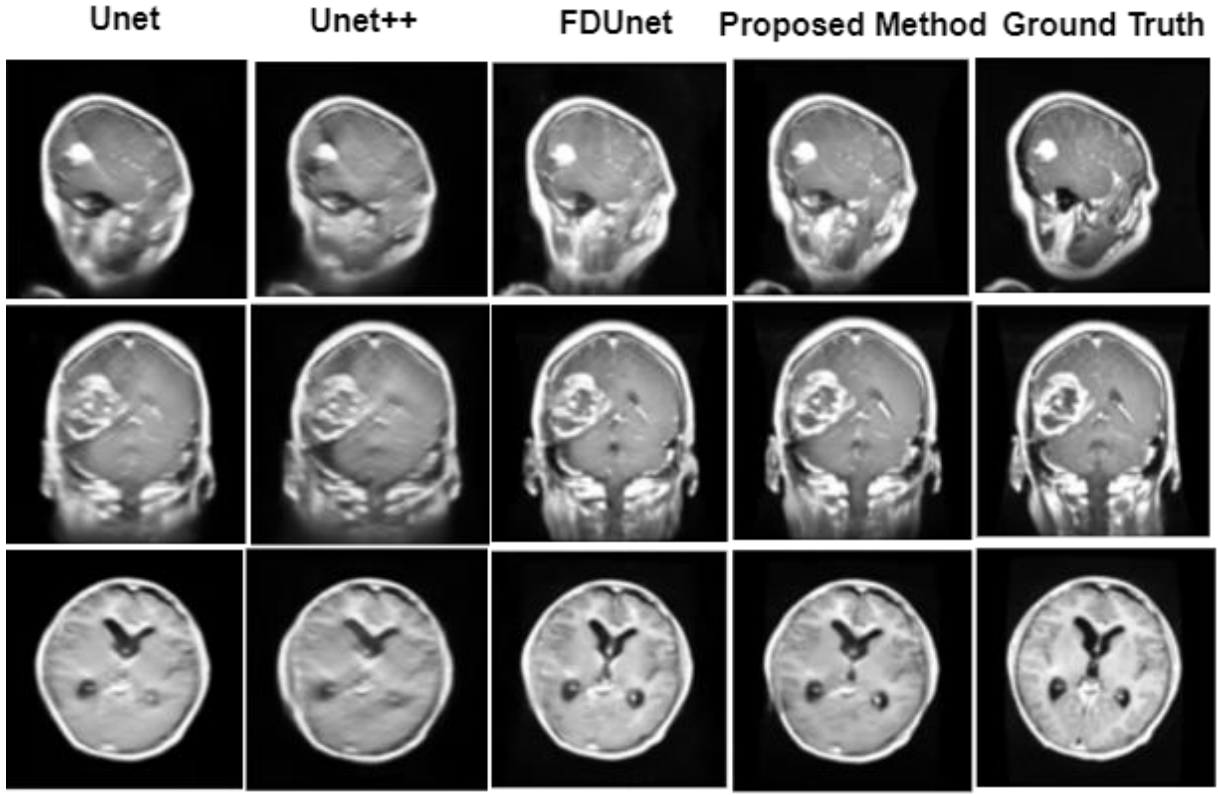

Fig 6. The effectiveness of DensePANet when used with MRI data related to brain tumors. The reconstructed images produced by the models UNet, UNet++, FD-UNet, and DensePANet are shown from left to right, together with the ground truth.

*4.4. Ablation Study*

In this part, we performed an ablation study to analyze the effect of each component of the proposed model on the performance of image reconstruction. It can be inferred from Table 3 that in the vessel and mouse experiments, our proposed method outperforms the other methods. However, in the brain experiment, vanilla GAN (Pix2Pix) has the highest SSIM and the wide UNet has the highest PSNR.

In fact, it is obvious that we can make considerable improvements on image reconstruction by employing GAN, regardless of the backbone on which GAN is built, i.e., UNet, UNet++, FD-UNet or FD-UNet++. It is seen that FD-UNet and FD-UNet++ result in the best outcome among above mentioned models. The results show that these two, effectively captures significant features to remove artifacts, and merging dense blocks enhances GAN performance by enabling the reuse of these features. In simpler words, including dense blocks not only tackle the problem of degradation but also strengthens the neural network and increases the quality of the reconstruction. Moreover, they extract comprehensive context information from input data, and then, the initial pressure distribution is obtained from a symmetric expanding path. Eventually, a high-quality artifact-free image is reconstructed.

Table 3 Quantitative results of our ablation study

| Dataset | Metric | UNet | FDUNet++ | GAN | GAN-UNet++ | Wide-UNet | GAN-FDUNet | GAN-FDUNet++ |
|---|---|---|---|---|---|---|---|---|
| Simulated Vessels | SSIM | 0.8308 ±0.1407 | 0.9271± 0.0010 | 0.9620± 0.0009 | 0.9656± 0.0043 | 0.9565± 0.0008 | 0.9585± 0.0091 | **0.9733± 0.0032** |
| | PSNR | 20.6166 ±2.9360 | 22.3993± 0.1224 | 24.1172± 0.1031 | 24.3852± 0.7431 | 23.7520± 0.0191 | 23.8454± 0.1408 | **25.7718± 0.5997** |
| Mouse-Abdomen | SSIM | 0.7509± 0.0154 | 0.8064 ± 0.0138 | 0.8284± 0.0005 | 0.8471± 0.0008 | 0.8190± 0.0002 | 0.8017 ± 0.0034 | **0.8602± 0.0064** |
| | PSNR | 17.7014± 0.4864 | 18.6775 ± 0.0117 | 18.8953± 0.0253 | 19.4226± 0.0603 | 19.2236± 0.0319 | 18.0156± 0.1515 | **20.0831± 0.2912** |
| Brain Tumor MRI | SSIM | 0.7523± 0.0242 | 0.8363± 0.0159 | **0.8637± 0.0012** | 0.8368± 0.0048 | 0.8515± 0.0034 | 0.8381± 0.0014 | 0.8504± 0.0011 |
| | PSNR | 20.5583± 0.4277 | 21.2580± 0.6000 | 22.1188± 0.0903 | 21.0641± 0.2657 | **22.1516± 0.0501** | 21.3528± 0.2742 | 22.0917± 0.0710 |

*4.5. Complexity Comparison*

Complexity analysis, also known as computational complexity analysis, is essential to evaluate the efficiency of algorithms in terms of the computational resources they require. In Table 4 we have compared complexity between different models used in this study, in terms of the number of model's parameters. It is observed that UNet has the lowest number of network parameters and exhibits the least effective reconstruction performance, as indicated in Tables 2 and 3. This can be attributed to the simplistic nature of its skip connections, leading to inadequate modeling of the changing relationships between image pairs within the basic encoder-decoder architecture. However, the UNet++ model addressed this limitation by incorporating dense nested skip connections, resulting in an increase in network parameters from 7.8 million to 9 million and enhancement in SSIM and PSNR values. Furthermore, the FDUNet approach achieves elevated SSIM and PSNR values, albeit with an increment in network parameters that is 66% higher than that of the UNet method. Finally, as evident in this table, our proposed method GAN-FDUNet++ (DensePANet) demonstrates superior results across two of three experiments, despite having approximately 17% fewer parameters compared to the GAN (pix2pix) model, the model with the greatest number of parameters or Wide UNet whose number of parameters is comparable with DensePANet works weaker than it in the all experiments except in brain MRI data PSNR's metric.

Table 4 Comparing complexity between different models

| Model | UNet | UNet++ | GAN-UNet++ | GAN-FDUNet++ | GAN (pix2pix) | GAN-FDUNet | FD-UNet | YNet | FDUNet++ | WUNet |
|---|---|---|---|---|---|---|---|---|---|---|
| **Parameters** | 7.8M | 9M | 13M | 22.8M | 26.7M | 16.9M | 13M | 10M | 18.8M | 21.7M |

## 5. Discussion

In this work, we did a post-processing based photoacoustic tomography image reconstruction to remove artifacts from an initial reconstruction caused by limited-view problem or more generally from sparse data. In fact, to achieve high-quality artifact-free reconstruction images, we proposed an improved GAN based model using the GAN's ability for image-to-image translation. In the context of artifact removal from reconstructed images, image-to-image translation means processing an artifactual image and obtain its artifact-free counterpart.

The outcomes of our experiments in this study indicate that the utilization of DensePANet contributes to the enhancement of image quality. While various methods may appear visually somewhat similar, the quantitative results reveal that DensePANet achieves considerably higher results in terms of SSIM and PSNR. For example, in mouse-abdomen data, we could achieve up to 1.15- and 1.17-fold improvement than UNet (UNet can be considered a common baseline for image reconstruction) model in SSIM, and PSNR, respectively. This improvement actually lies in the taking advantage of: 1) feature fusion of the UNet++, in other words, revised nested skip connections, implemented between convolution blocks, acquire feature maps at various semantic levels from the blocks while maintaining full resolution. The restructured skip pathways have the objective of minimizing the semantic gap between the feature maps of the encoder and decoder sub-networks. Consequently, the optimization process becomes more manageable when the feature maps from both networks exhibit greater semantic similarity. Also, we know the significance of low-level features in the reconstruction process. Nonetheless, integrating these low-level features with top-level features presents challenges due to the extended paths between them. Moreover, continuous convolution operations lead to the loss of low-order features. Ultimately, the prediction relying on the top will miss the chance to accumulate diverse information from both the bottom and the middle layers. 2) The incorporation of a fully dense block in every layer endows the network with density, enabling it to acquire additional feature maps. This augmentation in trainable parameters, from 7.8 to 13 million, empowers the FDUNet to assimilate more information. The use of dense blocks also addresses the vanishing gradient issue, ensuring that each feature map is densely connected to overcome the problem and encourage the reuse of information from preceding feature maps. This strategy not only prevents the learning of redundant features but also enhances the smooth flow of information within the network. 3) GAN-based networks rely on generative models and have the potential to generate images with a more natural appearance.

To strengthen our comparisons, we also implemented a customized wide UNet architecture with a comparable number of parameters to our proposed structure. This approach was taken to ascertain that the observed performance improvement in our architecture is not solely attributed to an increased number of parameters. It is also important to highlight that our suggested model exhibits low computational demands in terms of inference time. Specifically, reconstructing an image with dimensions of 256 * 256 pixels takes less than 0.08 seconds. Hence, our proposed method offers a promising solution for achieving real-time reconstruction once the model is adequately trained.

It is also important to highlight that, because of the utilization of nested skip connections between encoder and decoder branches, and dense blocks in both the encoder and decoder, the suggested reconstruction approach exhibits resilience to variations in object sizes and scales, spanning from small to large structures. Consequently, our reconstruction method has the capability to capture object reconstructions at multiple scales, proving particularly valuable for objects with abrupt changes in size and scale.

Nevertheless, there are a few potential constraints associated with the suggested approach.

First, a key constraint of this approach lies in the substantial number of parameters within the proposed architecture, totaling approximately 23 million. Consequently, a considerable volume of training samples is essential. Obtaining a sufficient quantity of ground truth (GT) data with high accuracy proves to be labor-intensive. To address the challenge of limited training samples, leveraging recently developed deep learning (DL) techniques is imperative for enhancing the performance of our network.

Furthermore, it's important to highlight that the experimental data for the mouse abdomen comprises a single tomographic projection setting, specifically 256 tomographic projections. To accommodate varying artifact intensities and properties in different settings, networks need to undergo re-training or re-tuning. Consequently, the parameters trained initially must be adapted to suit new datasets. Stability against data change, in order to enhance the robustness of the model, is still an open problem that we are interested in addressing in our future work.

## 6. Conclusion

In this study, we suggested and validated a new model, namely DensePANet, for accurate PAT image reconstruction from sparse data. We assessed our approach by employing three separate datasets and compared the experimental results to other works. Results prove that the DensePANet is superior to UNet and UNet++ and other solutions for image reconstruction using artifact removal. The advantage of the proposed method is mostly because of three reasons: first, generative models, being capable of augmenting as well as synthesizing new images that are very similar to real images, resulting in better performance in the task of image-to-image translation. Particularly, Pix2Pix generative adversarial networks can greatly decrease under-sampling artifacts and thus, improve reconstruction quality. Second, the inclusion of dense blocks in encoder and decoder branches greatly promotes the reuse of features and enhances the flow of information within the network. Additionally, this form of connectivity has a regularization effect that decreases the chances of overfitting. Third reason is the presence of nested skip connections between encoder and decoder parts which reduce the semantic gap between the feature maps of these sub-networks.

Overall, our model is capable of reducing artifacts of PAT, initiating from under-sampling and the limited-view problem. This method has the ability to significantly improve the quality of the image without making any changes to the imaging system or decreasing the speed of imaging as it is important for real-time imaging applications [39]. Furthermore, the model proposed in this study could potentially be applicable for reconstructing images using different types of tissues and a broader range of biomedical imaging techniques
including MRI and CT. In the future, our focus will be on examining more extensive and superior in vivo datasets, while also enhancing our technique to ensure greater results in photoacoustic imaging.

**References**

1-Tang, K., Zhang, S., Liang, Z.,Wang, Y., Ge, J., Chen,W., Qi, L., 2023. Advanced image post-processing methods for photoacoustic tomography: A review, in: Photonics, Multidisciplinary Digital Publishing Institute. p. 707.

2-Xu, M.,Wang, L.V., 2006. Photoacoustic imaging in biomedicine. Review of scientific instruments 77.

3-Zhou, Y., Yao, J.,Wang, L.V., 2016. Tutorial on photoacoustic tomography. Journal of biomedical optics 21, 061007–061007.

4-Wang, L.V., 2009. Multiscale photoacoustic microscopy and computed tomography. Nature Photonics 3, 503–509.

5-Lu,W., Huang, Q., Ku, G.,Wen, X., Zhou, M., Guzatov, D., Brecht, P., Su, R., Oraevsky, A., Wang, L.V., et al., 2010. Photoacoustic imaging of living mouse brain vasculature using hollow gold nanospheres. Biomaterials 31, 2617–2626.

6-Xia, J., Wang, L.V., 2013. Small-animal whole-body photoacoustic tomography: a review. IEEE Transactions on Biomedical Engineering 61, 1380–1389.

7-Lin, L., Hu, P., Shi, J., Appleton, C.M., Maslov, K., Li, L., Zhang, R.,Wang, L.V., 2018. Single-breath-hold photoacoustic computed tomography of the breast. Nature Communications 9, 1–9.

8-Burgholzer, P., Bauer-Marschallinger, J., Grün, H., Haltmeier, M., Paltauf, G., 2007. Temporal back-projection algorithms for photoacoustic tomography with integrating line detectors. Inverse Problems 23, S65.

9-Treeby, B.E., Cox, B.T., 2010. k-wave: Matlab toolbox for the simulation and reconstruction of photoacoustic wave fields. Journal of Biomedical Optics 15, 021314.

10-Szabo, T.L., 2004. Diagnostic ultrasound imaging: inside out. Academic press.

11-Guan, S., Khan, A.A., Sikdar, S., Chitnis, P.V., 2019. Fully dense unet for 2-d sparse photoacoustic tomography artifact removal. IEEE Journal of Biomedical and Health Informatics 24, 568–576.

12-Guan, S., Hsu, K.T., Eyassu, M., Chitnis, P.V., 2021. Dense dilated unet: deep learning for 3d photoacoustic tomography image reconstruction. arXiv preprint arXiv:2104.03130 .

13-Huang, C., Wang, K., Nie, L., Wang, L.V., Anastasio, M.A., 2013. Full wave iterative image reconstruction in photoacoustic tomography with acoustically inhomogeneous media. IEEE transactions on medical imaging 32, 1097–1110.

14-Ding, L., Razansky, D., Dean-Ben, X.L., 2020. Model-based reconstruction of large three-dimensional optoacoustic datasets. IEEE transactions on medical imaging 39, 2931–2940.

15-Hauptmann, A., Cox, B., 2020. Deep learning in photoacoustic tomography: current approaches and future directions. Journal of Biomedical Optics 25, 112903–112903.

16-Waibel, D., Gröhl, J., Isensee, F., Kirchner, T., Maier-Hein, K., Maier-Hein, L., 2018. Reconstruction of initial pressure from limited view photoacoustic images using deep learning, in: Photons Plus Ultrasound: Imaging and Sensing 2018, International Society for Optics and Photonics. p. 104942S.

17-Simonyan, K., Zisserman, A., 2014. Very deep convolutional networks for large-scale image recognition. arXiv preprint arXiv:1409.1556 .

18-Antholzer, S., Haltmeier, M., Schwab, J., 2019. Deep learning for photoacoustic tomography from sparse data. Inverse Problems in Science and Engineering 27, 987–1005.

19-Hauptmann, A., Lucka, F., Betcke, M., Huynh, N., Adler, J., Cox, B., Beard, P., Ourselin, S., Arridge, S., 2018. Model-based learning for accelerated, limited-view 3-d photoacoustic tomography. IEEE Transactions on Medical Imaging 37, 1382–1393.

20-Davoudi, N., Deán-Ben, X.L., Razansky, D., 2019. Deep learning optoacoustic tomography with sparse data. Nature Machine Intelligence 1, 453–460.

21-Vu, T., Li, M., Humayun, H., Zhou, Y., Yao, J., 2020. A generative adversarial network for artifact removal in photoacoustic computed tomography with a linear-array transducer. Experimental Biology and Medicine 245, 597–605.

22-Goodfellow, I., Pouget-Abadie, J., Mirza, M., Xu, B., Warde-Farley, D., Ozair, S., Courville, A., Bengio, Y., 2014. Generative adversarial nets. Advances in Neural Information Processing Systems 27.

23-Isola, P., Zhu, J.Y., Zhou, T., Efros, A.A., 2017. Image-to-image translation with conditional adversarial networks, in: Proceedings of The IEEE Conference on Computer Vision and Pattern Recognition, pp. 1125–1134.

24-Hakimnejad, H., Azimifar, Z., Nazemi, M.S., 2023. Unsupervised photoacoustic tomography image reconstruction from limited-view unpaired data using an improved cyclegan, in: 2023 28th International Computer Conference, Computer Society of Iran (CSICC), IEEE. pp. 1–6.

25-Krithika alias AnbuDevi, M., Suganthi, K., 2022. Review of semantic segmentation of medical images using modified architectures of unet. Diagnostics 12, 3064.

26-Salpea, N., Tzouveli, P., Kollias, D., 2022. Medical image segmentation: A review of modern architectures, in: European Conference on Computer Vision, Springer. pp. 691–708.

27-Ronneberger, O., Fischer, P., Brox, T., 2015. U-net: Convolutional networks for biomedical image segmentation. arxiv. Lecture Notes in Computer Science 2015.

28-Zhou, Z., Rahman Siddiquee, M.M., Tajbakhsh, N., Liang, J., 2018. Unet++: A nested u-net architecture for medical image segmentation, in: Deep Learning in Medical Image Analysis and Multimodal Learning For Clinical Decision Support. Springer, pp. 3–11.

29-Zhou, Z., Siddiquee, M.M.R., Tajbakhsh, N., Liang, J., 2019. Unet++: Redesigning skip connections to exploit multiscale features in image segmentation. IEEE Transactions on Medical Imaging 39, 1856–1867.

30-He, K., Sun, J., 2015. Convolutional neural networks at constrained time cost, in: Proceedings of The IEEE Conference on Computer Vision and Pattern Recognition, pp. 5353–5360.

31-Huang, G., Liu, Z., Van Der Maaten, L., Weinberger, K.Q., 2017. Densely connected convolutional networks, in: Proceedings of The IEEE Conference on Computer Vision and Pattern Recognition, pp. 4700–4708.

32-Mirza, M., Osindero, S., 2014. Conditional generative adversarial nets. arXiv preprint arXiv:1411.1784 .

33-Treeby, B.E., Cox, B.T., 2010. k-wave: Matlab toolbox for the simulation and reconstruction of photoacoustic wave fields. Journal of Biomedical Optics 15, 021314.

34-Staal, J., Abràmoff, M.D., Niemeijer, M., Viergever, M.A., Van Ginneken, B., 2004. Ridge-based vessel segmentation in color images of the retina. IEEE Transactions on Medical Imaging 23, 501–509.

35-Nickparvar, M., 2021. Brain tumor mri dataset. URL: https://www.kaggle.com/dsv/2645886, doi:10.34740/KAGGLE/DSV/2645886.

36-Wang, Z., Bovik, A.C., Sheikh, H.R., Simoncelli, E.P., 2004. Image quality assessment: from error visibility to structural similarity. IEEE Transactions on Image Processing 13, 600–612.

37-Kingma, D.P., Ba, J., 2014. Adam: A method for stochastic optimization. arXiv preprint arXiv:1412.6980 .

38- Lan, Hengrong, et al. "Y-Net: Hybrid deep learning image reconstruction for photoacoustic tomography in vivo." Photoacoustics 20 (2020): 100197.

39- Qin, Zezheng, et al. "Convolutional sparse coding for compressed sensing photoacoustic CT reconstruction with partially known support." Biomedical Optics Express 15.2 (2024): 524-539.